\documentclass[12pt]{iopjournal}

\usepackage{amsmath,amssymb}
\usepackage{booktabs}
\usepackage{graphicx}
\usepackage{siunitx}
\usepackage{xcolor}
\usepackage{flafter}
\usepackage{placeins}
\usepackage{hyperref}
\usepackage{orcidlink}

\renewcommand{\headrulewidth}{0pt}
\renewcommand{\footrulewidth}{0.5pt}
\fancyhfoffset[L]{0pt}
\renewcommand{\articletype}[1]{
  \noindent{\Large\sffamily Journal of Physics: Condensed Matter}\par
  \vspace{8mm}
  \noindent{\scriptsize\sffamily\bfseries\MakeUppercase{#1}}\par
  \vspace{5mm}
}

\newcommand{\Ghat}{\widehat{\Gamma}}
\newcommand{\rhat}{\widehat{r}}

\hypersetup{
  pdftitle={Observable dependence and rate identifiability in false-vacuum decay of the quantum Ising chain},
  pdfauthor={Boliang Yu, Ruixin Zhou, Hang Su},
  pdfsubject={Observable dependence and rate identifiability in false-vacuum decay of the quantum Ising chain},
  pdfkeywords={false-vacuum decay, quantum Ising chain, tensor networks, rate identifiability}
}

\begin{document}

\articletype{Paper}
\title{Observable dependence and rate identifiability\\
in false-vacuum decay of the quantum Ising chain}

\author{Boliang Yu$^{1,*}$\,\orcidlink{0009-0006-4664-5068}, 
Ruixin Zhou$^1$\,\orcidlink{0009-0004-9303-6674} 
and Hang Su$^1$\,\orcidlink{0009-0000-1686-5525}}

\affil{$^1$School of Physics and Astronomy, Shanghai Jiao Tong University,
800 Dongchuan Road, Shanghai 200240, China}

\par
\email{$^{*}$\href{mailto:yuboliang@sjtu.edu.cn}{yuboliang@sjtu.edu.cn}}

\begin{abstract}
Extracting a thermodynamic nucleation rate from finite-time quantum dynamics
requires separating observable decay from estimator and finite-size validity.
We develop a multilevel identification framework for real-time tensor-network
simulations of false-vacuum decay in the one-dimensional quantum Ising chain.
Across twelve parameter points, the same coherent two-kink amplitudes
semi-quantitatively predict both infinite-chain survival and magnetization
dynamics: the survival coefficient has a median lattice-to-theory ratio of
0.896, while the magnetization-area slope ratios span 0.767--0.931.  By
contrast, the microscopic nearest-neighbour bond response is coherence
dominated: vacuum--pair coherence contributes 60.0--81.5\% across seven points
with matched bond-dimension control, while substantial late-window slope
discrepancies remain that cannot be removed by a scalar normalization.  The
analysis establishes finite-time survival and magnetization benchmarks and
identifies the additional finite-size and branch-validation requirements for a
bulk thermodynamic rate interpretation.  Within the two-kink model and under
the adopted common normalization, the lattice-resolved WKB action gives a
median fixed-prefactor rate discrepancy of 4.13\% from the coherent-bubble
spectral calculation.  These results distinguish finite-time lattice--theory
consistency from the additional observable and finite-size evidence required
to identify a thermodynamic nucleation rate.
\end{abstract}


\section{Introduction}

Metastable decay is a canonical example of a process whose microscopic
dynamics and macroscopic description live at very different levels.  In
continuum quantum field theory the leading decay law is organized by the
bounce action and its fluctuation prefactor
\cite{Coleman1977,CallanColeman1977,Affleck1981}; related semiclassical
structures also underlie Schwinger pair production \cite{Schwinger1951}.
Statistical theories of nucleation and growth instead connect a local
nucleation process to a transformed fraction
\cite{JohnsonMehl1939,Barmak2010,Avrami1939,Avrami1940,Langer1967,Langer1969}.
These descriptions motivate an exponential survival law, but neither implies
that the coefficient of an exponential-looking fit to a finite quantum chain
is already a bulk nucleation rate.

The quantum Ising chain provides a controlled setting in which this issue can
be examined.  Its equilibrium solution and scaling limit are well understood
\cite{Pfeuty1970,McCoyWu1978}, and confinement and weak-longitudinal-field
asymptotics have been studied analytically
\cite{Rutkevich1999,FonsecaZamolodchikov2003}.  Rutkevich derived the weak-field
decay rate of the metastable chain, including lattice-spectrum effects.
Maertens \emph{et al.}\ subsequently used real-time MPS and a bubble-state
description to resolve nucleation, growth, and Bloch-limited regimes and to
compare an intermediate-time rate directly with that prediction.  Johansen
\emph{et al.}\ developed coherent bubble theory in the thermodynamic chain and
benchmarked bubble density, magnetization, and spin correlations against MPS
through intermediate times \cite{Maertens2025,Johansen2025}.  These studies
establish positive intermediate-time benchmarks.  Here we independently
quantify that coherent two-kink baseline on a uniform parameter grid, extend the
same amplitudes to the microscopic nearest-neighbour bond operator, and test
whether the available finite-size evidence supports a thermodynamic
extrapolation.  Real-time lattice studies have also exposed false-vacuum decay,
confinement, Bloch oscillations, and finite-size effects using exact
diagonalization and tensor-network methods
\cite{Kormos2017,Verdel2020,Lagnese2021,Pomponio2022,Lerose2020}.
Recent theoretical and numerical work has further developed the many-body
description of metastability \cite{Maki2023,Batini2024,Yin2025}.  Platform-oriented
and analogue-simulation studies \cite{Darbha2024,Zhu2024,Vodeb2025}, and
experimental studies \cite{Zenesini2024,Chao2026}, have made the question
increasingly operational \cite{Milsted2022,Karpov2022,Vovrosh2022,Ge2026}.  In
particular, coherent finite-size
oscillations can replace apparently irreversible decay at resonances
\cite{Ge2026}, while field-theory studies show that the exponential dependence
need not fix a model-independent overall coefficient \cite{Lencses2022}.

The central challenge is to separate agreement over an accessible time window
from a thermodynamic rate interpretation.  A finite trace can admit a
high-\(R^2\) linear fit after a logarithmic transformation while the fitted
coefficient still depends on the readout, fit interval, boundary condition,
system size, or tensor-network cutoff.  These dependencies probe distinct
physical questions, so we distinguish finite-time lattice observables,
coherent two-kink theory, and semiclassical action comparisons.

We make three contributions.  First, we test a single coherent two-kink
construction across survival, magnetization, and the nearest-neighbour bond.
The macroscopic observables are captured semi-quantitatively, whereas the bond
response is dominated by vacuum--pair coherence and exhibits a late-window
mismatch that cannot be removed by scalar normalization.  Second, we identify
the preparation, finite-time, and finite-size requirements that prevent the
available data from supporting a bulk-rate extrapolation.  Third, within the
same two-kink description, we compare lattice-resolved and continuum
semiclassical actions, separating normalization-dependent rate offsets from
normalization-independent parameter dependence.  These comparisons separate
finite-time lattice--theory consistency from the additional finite-size evidence
required for a bulk-rate interpretation.

\section{Model, operational target, and relevant scales}

\subsection{Quantum Ising chain}

We consider a ferromagnetic quantum Ising chain with transverse and
longitudinal fields,
\begin{equation}
 H=-J\left[\sum_{j}\sigma^x_j\sigma^x_{j+1}
   +\sum_j\left(h_\perp\sigma^z_j+h_\parallel\sigma^x_j\right)\right],
\label{eq:H}
\end{equation}
with either open (OBC) or periodic (PBC) boundary conditions.  The intended
 initial state lies in the positive-magnetization sector that becomes
 metastable after the longitudinal-field sign is reversed.  Whether an existing
 finite-system state actually lies on that branch is treated below as a
 separate validity question.  All quoted times and rates are in units set by $J=\hbar=a=1$.  The
 post-quench field obeys $h_\parallel<0$, while tables and figures report its
 magnitude.  We use $\rhat_{i,g}$ for a fitted local slope of readout $i$ at
 parameter point $g$, and $\Gamma$ for a planted, theoretical, or operational
 rate.  The finite-time lattice-observable analysis, coherent-bubble-theory
 (CBT) spectral calculation, and semiclassical action comparison are denoted
 L0, L1, and L2, respectively.  The parameter set contains a principal $4\times4$ grid,
$h_\perp\in\{0.70,0.75,0.80,0.85\}$ and $S_0\in\{3,4,5,6\}$, together with
six edge probes at $h_\perp=0.60,0.90$ and $S_0=3,4,5$, for 22 points in
 total.  Seventeen points have non-empty admissible intervals.  All sixteen
 principal-grid points have non-empty baseline iMPS intervals.  For the
cross-level analysis reported here, a uniform feasibility screen was applied
to all sixteen principal-grid points and selected the twelve points with
$S_0=3,4,5$.  The screen required
$\mathcal M_{\rm lat}^{(256)}=256e^{-S_{\rm lat}}/(2\pi)\geq0.1$, where
$S_{\rm lat}$ is the lattice-resolved action defined below, together with
$\mathcal T>1$, an estimated required time no greater than $100/J$, and
$S_0\geq2$.  Nine points span the full physical window, whereas the three
$h_\perp=0.85$ intervals end at $t_\chi$.  The four $S_0=6$ points fail the
signal condition; they consequently received only the baseline numerical
setting and are retained as descriptive weak-signal checks.  A
separate $20\times20$ CBT scan over $h_\perp\in[0.50,0.98]$ and
  $S_0\in[2,12]$ tests the L1 finite-window estimator and maps the physical-window and L1--L2 validity region.

The thermodynamic spontaneous magnetization, kink mass, exact zero-field kink dispersion,
and post-quench string tension are
\begin{align}
 M_{\rm th}&=(1-h_\perp^2)^{1/8}, & m&=2J(1-h_\perp),\\
 \epsilon(\theta)&=2J\sqrt{1-2h_\perp\cos\theta+h_\perp^2},
 & f&=2J|h_\parallel|M_{\rm th} .
\label{eq:microscopic}
\end{align}
The parameter grid is parameterized by the leading continuum target action
\begin{equation}
 S_0=\frac{\pi(1-h_\perp)^2}{|h_\parallel|M_{\rm th}},\qquad
 |h_\parallel|=\frac{\pi(1-h_\perp)^2}{S_0M_{\rm th}}.
\label{eq:s0}
\end{equation}
This parameterization holds the nominal exponential difficulty fixed while
$h_\perp$ changes the lattice corrections and accessible dynamical scales.

\begin{figure}[!htbp]
\centering
\includegraphics[width=0.96\linewidth]{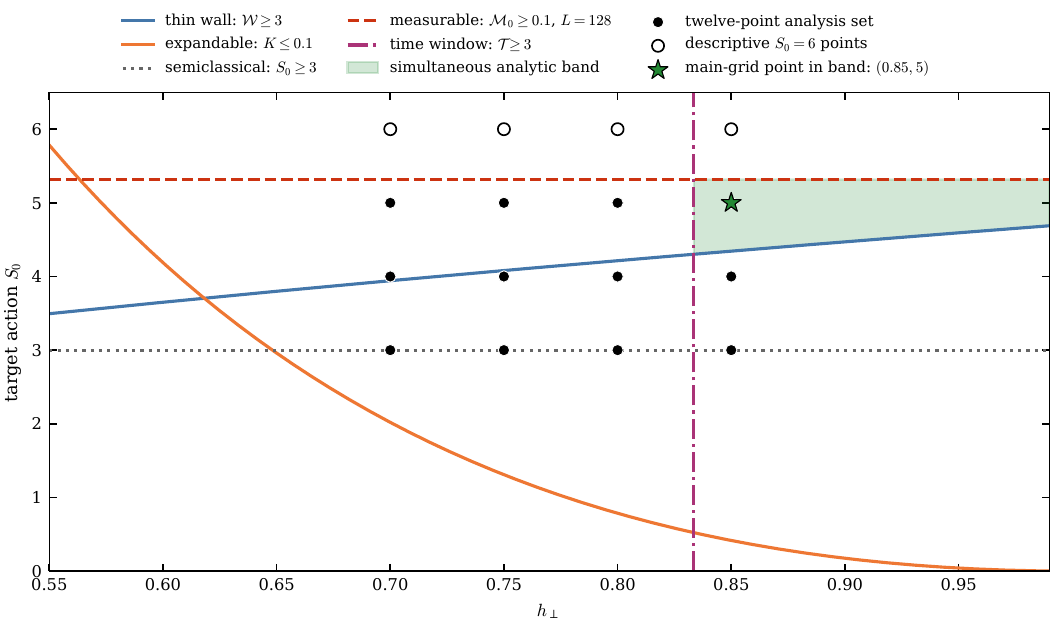}
\caption{Analytic scale-compatibility map.  The horizontal axis is the transverse field $h_\perp$ and the
vertical axis is the target continuum action $S_0$; both quantities are
dimensionless.  The blue, orange, dotted black, dashed red, and dash-dotted purple
boundaries impose $\mathcal W\geq3$, $K\leq0.1$, $S_0\geq3$,
$\mathcal M_0\geq0.1$ at $L=128$, and $\mathcal T\geq3$, respectively; the
green area is the band satisfying all five analytic conditions.  Filled circles
show the twelve-point analysis set, open circles the four descriptive $S_0=6$
points, and the star marks the sole principal-grid point in the band,
$(h_\perp,S_0)=(0.85,5)$.  Throughout
the figures $J=\hbar=a=1$, so energies are measured in $J$ and times in
$J^{-1}$.
Initial-state, numerical, estimator-specific, size, geometry, and cross-engine
checks are applied separately and are not shown here.}
\label{fig:constraints}
\end{figure}

The resonant bubble length, upper classical turning point, and thin-wall width
estimate are
\begin{equation}
 l_r=\frac{2m}{f}=\frac{2(1-h_\perp)}{|h_\parallel|M_{\rm th}},
 \qquad l_r^+=\frac{2(1+h_\perp)}{|h_\parallel|M_{\rm th}},
 \qquad W\simeq\frac{\sqrt{h_\perp}}{1-h_\perp}.
\label{eq:lengths}
\end{equation}
The three physical time scales used below are defined operationally by
\begin{equation}
 T_\Omega=\frac{1-h_\perp}{J|h_\parallel|M_{\rm th}},\qquad
 T_\Delta=\frac{1}{f},\qquad
 T_{\mathrm{Bloch}}=\frac{2\pi}{f}.
\label{eq:timescales}
\end{equation}
Four dimensionless controls expose the competing requirements,
\begin{equation}
 K=\frac{f}{4Jh_\perp},\quad
 \mathcal W=\frac{l_r}{W},\quad
 \mathcal T=\frac{T_\Delta}{T_\Omega}=\frac{1}{2(1-h_\perp)},\quad
 \mathcal M_0=\frac{Le^{-S_0}}{2\pi}.
\label{eq:controls}
\end{equation}
Small $K$ favours an expandable bubble, large $\mathcal W$ a controlled thin
 wall, large $\mathcal T$ a longer golden-rule interval, and sufficiently
 large $\mathcal M_0$ a measurable finite-size signal.  These controls cannot
 all be improved simultaneously; figure~\ref{fig:constraints} maps their
 competition.
The calligraphic signal control $\mathcal M_0$ is the $L=128$ continuum-action
proxy used in the analytic scale map.  It is distinct from both the
lattice-action quantity $\mathcal M_{\rm lat}^{(256)}$ used in the feasibility
screen and the parameter-point-specific magnetization normalizer $M_{0,g}$
introduced below.
The feasibility screen gives minimum numerical-analysis eligibility; it does
not assert analytic control or semiclassical validity.  By contrast, the five
conditions in figure~\ref{fig:constraints}, including $\mathcal T\geq3$, define
a stricter analytic scale-compatibility region.

\FloatBarrier

The green band in figure~\ref{fig:constraints} represents analytic scale
compatibility, not overall numerical optimality.  On the present grid, moving
toward it enlarges the characteristic bubble and brings an earlier
bond-dimension limitation, shortening the accessible real-time window.  The sole
principal-grid point satisfying all five analytic conditions,
$(h_\perp,S_0)=(0.85,5)$, is included in the twelve-point comparison and, as
shown below, remains consistent with the survival and magnetization two-kink
benchmarks, but its usable interval ends at the bond-dimension convergence time
$t_\chi$ defined below.  We therefore retain the planned reference point
$(0.75,4)$, which spans the full physical window, as the representative
trajectory rather than replacing it with the scale-compatible but numerically
shorter $(0.85,5)$ case.

The operational target is not simply the slope of a transformed trace.
For a system of length \(L\), a readout-specific fitted coefficient is denoted
\(\rhat_i(L)\).  Under OBC we test the intensive boundary decomposition
\begin{equation}
 \rhat_i^{\mathrm{OBC}}(L)=r_{i,\mathrm{bulk}}
          +\frac{2r_{i,\mathrm{edge}}}{L},
\label{eq:obc}
\end{equation}
whereas under PBC we test the working extrapolation model
\begin{equation}
 \rhat_i^{\mathrm{PBC}}(L)=r_{i,\mathrm{bulk}}+\frac{a_i}{L}.
\label{eq:pbc}
\end{equation}
Unlike the OBC endpoint term, the PBC $1/L$ correction is not derived here.
The available branch-screened data do not permit comparison with $1/L^2$ or
exponential alternatives, so equation~(\ref{eq:pbc}) is not treated as a
validated finite-size law.  The planted recovery test below uses the same
$1/L$ form and therefore tests numerical recovery within that model, not the
model's applicability to the MPS data.

Only a coefficient that is locally stable, numerically converged, and
consistent with this geometry test is interpreted as an operational
\(\Ghat_i=r_{i,\mathrm{bulk}}\).  Agreement with a continuum or semiclassical
prediction is a subsequent test, not part of the definition.
For scale context, the finite-size coverage summary compares every available $L$ with the
resonant length $l_r$, the upper classical turning point $l_r^+$, and the
centre-to-boundary time $L/(2v_{\max})$, where
$v_{\max}=\max_\theta|\partial_\theta\epsilon(\theta)|=2Jh_\perp$.
The simulated size sets are OBC $L=32,64,128,256$ and PBC $L=24,32,48$.
Of 154 nominal point--geometry--size combinations, 65 have a baseline physical upper
edge beyond this boundary-arrival time.  Among the 44 smallest-size OBC/PBC
cases, 3 do not contain $l_r$ and 40 do not contain $l_r^+$.  These ratios are
reported as applicability diagnostics rather than additional exclusion criteria;
in particular, the 65/154 count is not a count of final estimator-specific fits.

\subsection{Physical and numerical time scales}

Several time scales constrain the interval in which a nucleation
interpretation can be attempted.  The lower edge is
\begin{equation}
 t_{\mathrm{lo}}=\max\left(T_\Omega,\frac{2}{m},t_{\mathrm{settle}}\right),
\label{eq:tlo}
\end{equation}
where \(T_\Omega\) resolves the driving or oscillation scale, \(m^{-1}\)
resolves the massive microscopic response, and
\(t_{\mathrm{settle}}=0.5/J\) excludes the preparation transient.  The base
physical upper edge is
\begin{equation}
 t_{\mathrm{hi}}^{\mathrm{base}}=
 \min\left(T_\Delta,\frac{T_{\mathrm{Bloch}}}{4}\right).
\label{eq:thibase}
\end{equation}
The quarter-Bloch bound is retained to expose its physical origin, although
$T_{\mathrm{Bloch}}/4=(\pi/2)T_\Delta$ makes it inactive for the present
definitions.  The trajectory-specific MPS upper edge for $R_1$, $R_2$, and $R_4$ is
\begin{equation}
 t_{\mathrm{hi}}^{\mathrm{MPS}}=
 \min\left(t_{\mathrm{hi}}^{\mathrm{base}},t_{\mathrm{final}},t_\chi\right),
\label{eq:thimps}
\end{equation}
where $t_{\mathrm{final}}$ is the final simulated time and $t_\chi$ is
the last sampled time before the first maximum absolute
difference between the $\chi=128$ and $\chi=64$ trajectories reaches $10^{-3}$.
The maximum is taken over magnetization, normalized survival,
domain-wall excess, and the available connected correlators; it is not a generic
relative-error tolerance.  The correlation-derived readout $R_3$ additionally uses
\begin{equation}
 t_{\mathrm{hi}}^{R_3}=\min\left(t_{\mathrm{hi}}^{\mathrm{MPS}},t_{\mathrm{tail}}\right),
\qquad
\frac{|R_3(t_{\mathrm{tail}})-R_4(t_{\mathrm{tail}})|}
{|R_3(t_{\mathrm{tail}})|}>0.05
\label{eq:tail}
\end{equation}
at the first threshold crossing.  Algebraically, this criterion tests correlation-tail closure,
normalization, and averaging consistency; it is not an independent
bubble-overlap time.  We therefore apply it only to $R_3$, whose definition
requires the available correlation tail, and use the base MPS bound for $R_1$,
$R_2$, and $R_4$ in every geometry.  In the synthetic recovery calculation
both MPS limits are set to infinity.

The order-one factors in equations~(\ref{eq:tlo})--(\ref{eq:thibase}) are
tested by varying the mass bound from $1/m$ to $3/m$, the settling time from
$0.25/J$ to $1/J$, and the upper bound from $0.75T_\Delta$ to
$1.25T_\Delta$, one at a time.  Whether the reference point satisfies the plateau criteria does not
change.  The largest R1 selected/full-window shifts are 4.25\%/3.96\%; the
27.98\% largest full-window shift belongs to an $R_4$ diagnostic that already
does not satisfy the local-plateau criterion.  Complete branch-screened PBC and
three-geometry coverage remain zero for every tested window.

\section{Readouts and analysis protocol}

\subsection{Four readouts and three structural channels}

For finite systems define the fidelity per site
\begin{equation}
s(t)=|\langle\psi(0)|\psi(t)\rangle|^{2/L},
\qquad R_1(t)=-\ln[s(t)/s(0)],
\label{eq:r1}
\end{equation}
with the corresponding transfer-matrix fidelity for iMPS.  The magnetization
channel uses the simulated initial magnetization
$M_{0,g}\equiv m_{x,g}(0)$ for each parameter point $g$ and the fixed-velocity KJMA
transformation
\begin{equation}
-\ln\left[\frac{m_x(t)/M_{0,g}+1}{2}\right]
=\beta t^2+\mathrm{const},\qquad
\rhat_2=\frac{\beta}{v},\quad v=2Jh_\perp .
\label{eq:r2}
\end{equation}
Only $\beta=\Gamma v$ is structurally identified by $R_2$; fixing $v$
makes it a full-window cross-check, and the analysis excludes it from
formal finite-size geometry inference.

For the connected longitudinal correlator
$C_r=\langle\sigma_i^x\sigma_{i+r}^x\rangle-
\langle\sigma_i^x\rangle\langle\sigma_{i+r}^x\rangle$, $R_3$ constructs
\begin{equation}
P_n(t)=\frac{C_{n-1}(t)-2C_n(t)+C_{n+1}(t)}{4M_{0,g}^2},
\qquad Q_3(t)=\sum_{n=1}^{N}P_n(t),\quad R_3(t)=Q_3(t)-Q_3(0).
\label{eq:r3}
\end{equation}
The analysis uses the maximum available cutoff $N=48$;
$N=36$ is used only in the post-hoc correlation-cutoff sensitivity diagnostic.
The fourth readout is the excess bond-defect density,
\begin{equation}
d_{\mathrm{dw}}(t)=
\frac{1-\langle\sigma_i^x\sigma_{i+1}^x\rangle_t}{2},
\qquad R_4(t)=\frac{d_{\mathrm{dw}}(t)-d_{\mathrm{dw}}(0)}{2M_{0,g}^2}.
\label{eq:r4}
\end{equation}
The discrete identity
\begin{equation}
\sum_{n=1}^{N}P_n=\frac{C_0-C_1-C_N+C_{N+1}}{4M_{0,g}^2}
\label{eq:r34}
\end{equation}
makes $R_3$ and $R_4$ conditionally equivalent after the correlation tail
has saturated and the local-bond conventions match.  They are therefore an
order-parameter-normalized tail-consistency pair, not independent rate
evidence.  The factors of $M_{0,g}^2$ define that coarse-grained normalization;
they are not asserted to be exact microscopic quasiparticle weights.  The
three structural channels are survival ($R_1$), fixed-velocity magnetization
($R_2$), and the algebraically linked correlation/bond diagnostics
($R_3$--$R_4$).
Each readout is transformed so that its ideal late-transient, pre-collision
behaviour is linear.  Fits are performed only inside the intersection of the
readout domain and the admissible time interval.  A simulation can therefore
exist without producing a valid transformed coefficient.  This occurs for \(R_2\):
all four OBC sizes were simulated at all sixteen grid points, but only eight
points satisfy the transformation-domain and subsequent analysis criteria.
At all eight high-field points in the principal grid, $h_\perp=0.80,0.85$ and
$S_0=3,4,5,6$, the transformed quantity at \(L=32\) is undefined; those
cases lie outside the transformation domain rather than reflecting missing simulations.

\subsection{Microscopic bond operator in the coherent two-kink sector}

The normalized diagnostic $R_4$ should be distinguished from a microscopic
prediction for the raw bond observable.  Write the physical transverse field
as $g=Jh_\perp$.  In a positive-momentum Nambu block, differentiating the
pre-quench Hamiltonian with respect to $J$ at fixed $g$ gives the total bond
operator.  Rotation to the Bogoliubov basis yields
\begin{equation}
 w(\theta)=\frac{1-h_\perp\cos\theta}
 {\sqrt{1+h_\perp^2-2h_\perp\cos\theta}},\qquad
 q(\theta)=\frac{h_\perp\sin\theta}
 {\sqrt{1+h_\perp^2-2h_\perp\cos\theta}},
\label{eq:bondsymbols}
\end{equation}
where $-4w$ is the pair-sector change of the total bond and $2\mathrm{i}q$
is its vacuum--pair matrix element.  In the positive-separation basis define
\begin{align}
 W_{nm}&=\frac{2}{\pi}\int_0^\pi w(\theta)
 \sin(n\theta)\sin(m\theta)\,{\rm d}\theta,\nonumber\\
 b_n&=\frac{2}{\pi}\int_0^\pi q(\theta)\sin(n\theta)\,{\rm d}\theta.
\label{eq:bondkernels}
\end{align}
Let $c_L(t)\equiv\alpha_L(t)/\sqrt{N_s}$ be the intensive coherent mode
amplitude, given explicitly in equation~\eqref{eq:cbtfinite}, and define
$a_n(t)=\sum_L\phi_L(n)c_L(t)$.  The vacuum--two-kink projection of the
physical raw bubble proxy is
\begin{equation}
 \mathcal D_{\rm bond}^{(2)}(t)
 \equiv\frac{d_{\rm dw}(t)-d_{\rm dw}(0)}{2}
 =-\frac12\operatorname{Re}[\boldsymbol b^{\mathsf T}\boldsymbol a(t)]
 +\operatorname{Re}[\boldsymbol a^\dagger(t)\boldsymbol W\boldsymbol a(t)].
\label{eq:bondprojection}
\end{equation}
Here the first term is vacuum--pair coherence.  The second is the pair--pair
term: it is diagonal in momentum space but generally non-diagonal in the
relative-coordinate basis.  At $h_\perp=0$, $\boldsymbol b=0$ and
$\boldsymbol W=\boldsymbol 1$, so equation~\eqref{eq:bondprojection} reduces
to the two-kink population per site $\sum_L|c_L(t)|^2$, later denoted by
$\mathcal N(t)$ in equation~\eqref{eq:cbtfinite}.
The projection is exact for the pre-quench free-fermion bond operator within
the retained vacuum and zero-total-momentum two-kink sectors, with the
coherent expectation kept through quadratic order.  It omits higher-kink
sectors, inter-bubble corrections, and longitudinal-field interaction
corrections.  The derivation and numerical symbol checks are given in the
supplementary material.

\subsection{State preparation, evolution, and validation}

The false-vacuum branch is the positive-$x$ ordered state, and the negative
post-quench longitudinal field favours the opposite polarization.  Infinite
MPS preparation uses the positive-$x$ product state only to initialize
imaginary-time iTEBD at $h_\parallel=0$; the optimized symmetry-broken iMPS,
not the product state, is the real-time quench state.  Finite OBC and PBC
states are optimized by DMRG with a positive selecting field
$\delta=10^{-6}$ and are then switched directly in one quench to
$h_\parallel^{\rm post}<0$, without an intermediate field-removal or ramp
stage.  This prescription states the intended preparation but does not prove
that a short finite system has selected the intended polarized branch.

For the available finite-system data we therefore apply the conservative
branch-selection criterion $m_x(0)>0$ and
$|M_{0,g}/M_{\rm th}-1|\leq\epsilon$.  Here $M_{\rm th}$ is reserved for
the thermodynamic branch screen, whereas $M_{0,g}$ is the
normalizer used in equations~(\ref{eq:r2})--(\ref{eq:r4}).  This is an
operational data-sufficiency test, not a universal or sufficient definition
of finite-size branch validity; finite size, boundary conditions, selecting
field, and bond dimension can all shift $m_x$ from the thermodynamic zero-field
value.  The branch-selection screen is independent of subsequent dynamical
behavior and tests whether the prepared state belongs to the intended
metastable branch before time evolution.  All twelve analysis-set iMPS points
satisfy the criterion at one percent.  The complete OBC count changes from zero at one
and two percent to three at five and ten percent, whereas the complete PBC and
three-geometry counts remain zero throughout the tested one-to-ten-percent
range.  The available $\delta=10^{-4},10^{-5},10^{-6}$ sensitivity
results apply only to selected controls and do not validate every available
finite-chain initial state.

At the representative preparation point, uniform selecting fields yield
stable static intervals for OBC $L=64$ and PBC $L=32,48$, but not for OBC
$L=32$ or PBC $L=24$; endpoint pinning stabilizes the local OBC branch but was
not removed before real-time evolution.

Infinite-chain, OBC, and PBC dynamics were obtained with two-site iTEBD,
finite-system TEBD, and TDVP, respectively
\cite{Vidal2003,Vidal2004,Daley2004,WhiteFeiguin2004,Haegeman2011,Haegeman2016}.
We compared $\chi=64$ and 128 and performed time-step, exact-diagonalization,
and free-fermion cross-checks following established MPS practice
\cite{White1992,Schollwock2011,Paeckel2019}.  Survival and magnetization are
stable at the level relevant to the reported benchmark, whereas the
correlation/bond channels show substantially larger bond-dimension sensitivity;
full settings and validation results are given in the supplementary material.
\subsection{Window scan and plateau definition}

Within \([t_{\mathrm{lo}},t_{\mathrm{hi}}]\), the scan window has width
\(0.20(t_{\mathrm{hi}}-t_{\mathrm{lo}})\), and adjacent starting points are
separated by one quarter of that width.  For every valid window we record its
linear-fit slope and fit error.  Consecutive groups containing at least three
windows qualify as a plateau when
\begin{equation}
 \delta_{\mathrm{plat}}=
 \frac{\max_k|s_k-\bar{s}|}{|\bar{s}|}<0.10.
\label{eq:plateau}
\end{equation}
The longest qualifying group is selected; ties are resolved by preferring the
smaller relative spread and then the later group.  Its mean slope is reported
as \(\rhat_{i,g}\).  The within-window fit component is the root-mean-square
of the individual fit errors, while the window component is half the range of
the accepted slopes.  If no group qualifies, the diagnostic output is the
median of all window slopes; because the plateau criterion is not satisfied,
this diagnostic is not interpreted as a rate.

\subsection{Extraction protocol and retrospective checks}

We distinguish finite-time coefficients from a bulk-rate interpretation by
requiring valid state preparation and evolution, an admissible transformed
fitting window with stable local slope, and complete finite-size coverage with
consistent geometry extrapolations.  Retrospective initial-state,
correlation-cutoff, and window-sensitivity checks qualify individual channels
but do not restore a periodic size series prepared in the intended metastable
branch; details are given in the supplementary material.

Reference-point readout coefficients and continuous traces are reported in the
supplementary material; the finite-size coverage is summarized once with the
results below.

\FloatBarrier
\section{Kinetic and semiclassical descriptions}

\subsection{Lattice-resolved tunnelling action}

Following the lattice semiclassical construction of
Refs.~\cite{Rutkevich1999,Maertens2025}, consider a zero-total-momentum kink
pair with positive relative coordinate $x$.  The semiclassical two-body
Hamiltonian is
\begin{equation}
 H_{\mathrm{rel}}(x,\theta)=2\epsilon(\theta)-fx.
\label{eq:hrel}
\end{equation}
The classically forbidden branch has $\theta=i\kappa$.  Continuing the exact
lattice dispersion in equation~(\ref{eq:microscopic}) gives the
lattice-resolved WKB action
\begin{equation}
 S_{\mathrm{lat}}=\frac{8J}{f}
 \int_0^{\ln(1/h_\perp)}
 \sqrt{1-2h_\perp\cosh\kappa+h_\perp^2}\,d\kappa .
\label{eq:slat}
\end{equation}
We compare it with two continuum forms,
\begin{equation}
 S_{\mathrm{cont}}^{(0)}=S_0,
 \qquad S_{\mathrm{cont}}^{(1)}=\frac{S_0}{\sqrt{h_\perp}}.
\label{eq:scont}
\end{equation}
Writing $h_\perp=e^{-\eta}$ makes the near-critical relation explicit:
\begin{align}
 S_{\mathrm{lat}}&=\frac{2\pi J}{f}\eta^2e^{-\eta/2}
 \left(1+\frac{5\eta^2}{96}+O(\eta^4)\right),\\
 \frac{S_{\mathrm{lat}}}{S_{\mathrm{cont}}^{(1)}}
 &=1-\frac{\eta^2}{32}+O(\eta^4).
\label{eq:slatexpansion}
\end{align}
The comparison holds the prefactor construction fixed,
\begin{equation}
 \Gamma_k=\frac{f}{2\pi}e^{-S_k},\qquad
 k\in\{\mathrm{cont0},\mathrm{cont1},\mathrm{lat}\}.
\label{eq:fixedprefactor}
\end{equation}
Here $f/(2\pi)$ is a common reference prefactor used to compare the action
choices; it is not the full Rutkevich prefactor.  Consequently, differences
among the three curves isolate the rate discrepancy induced by the action
choice rather than determining a complete arbitrary-field lattice prefactor.

\subsection{Coherent bubble theory and the spectral rate}

Building on the confined-kink and coherent-bubble formulations of
Refs.~\cite{Rutkevich1999,FonsecaZamolodchikov2003,Johansen2025}, we give a
self-contained L1 construction in the normalization and finite-basis
conventions used throughout this work.  The kink--antikink relative coordinate
obeys the antisymmetrized Wannier--Stark problem
\begin{equation}
 H^{(2)}_{nn'}=T_{n-n'}-T_{n+n'}-fn\delta_{nn'},
\label{eq:cbthamiltonian}
\end{equation}
on the positive-separation basis $n=1,\ldots,n_{\max}$.  The antisymmetric
image term imposes the hard wall at $n=0$.  The hopping coefficients are the
discrete Fourier components of the exact two-kink dispersion,
\begin{equation}
 T_d=\frac{2}{N_\theta}\sum_{j=0}^{N_\theta-1}
 \epsilon(2\pi j/N_\theta)e^{-2\pi i j d/N_\theta},
 \qquad N_\theta=\max(32768,4n_{\max}),
\label{eq:cbthopping}
\end{equation}
with the real coefficient retained for $d=0,\ldots,2n_{\max}$.  If
$\phi_L(n)$ is a normalized eigenmode, its intensive coupling to the
metastable state is
\begin{equation}
 \bar\Omega_L=\frac12 M_{\rm th}J|h_\parallel|
 \sum_{n\ge1}h_\perp^n\phi_L(n).
\label{eq:source}
\end{equation}
The calculation linearly interpolates $|\bar\Omega_L|^2$ between the adjacent
ordered spectral levels that bracket $E=0$.  Since the local level density is
$\rho(0)=1/f$, the spectral Fermi-golden-rule (FGR) rate per site is
\begin{equation}
 \Gamma_{\mathrm{CBT,FGR}}
 =2\pi\rho(0)|\bar\Omega(0)|^2
 =\frac{2\pi}{f}|\bar\Omega(0)|^2.
\label{eq:cbtfgr}
\end{equation}
No fitted normalization, artificial broadening, or linewidth parameter is
introduced.  Spectra satisfying the turning-point cutoff are retained; basis
construction and convergence checks are given in the supplementary material.
The relative fixed-prefactor discrepancy reported below is
\begin{equation}
 \varepsilon_k=
 \left|\frac{\Gamma_{\mathrm{CBT,FGR}}}{\Gamma_k}-1\right|.
\label{eq:epsilon}
\end{equation}
CBT and the lattice WKB calculation share the exact lattice kink dispersion
and a coherent two-kink reduction.  Their comparison is therefore an internal
benchmark of the rate discrepancy induced by the action choice under shared
assumptions.

The finite-time CBT signal follows directly from first-order excitation of
the discrete modes.  For time-independent coupling after the quench,
\begin{equation}
 c_L(t)=\bar\Omega_L\frac{e^{-iE_Lt}-1}{E_L},\qquad
 \mathcal N(t)=\sum_L\frac{4|\bar\Omega_L|^2}{E_L^2}
 \sin^2\left(\frac{E_Lt}{2}\right).
\label{eq:cbtfinite}
\end{equation}
Within this dilute, noninteracting coherent-bubble construction, the survival
probability and fidelity-per-site transform are
\begin{equation*}
 P_{\rm surv}^{\rm CBT}(t)
 =|\langle 0_{\rm FV}|\Psi_{\rm CBT}(t)\rangle|^2
 =e^{-N_s\mathcal N(t)},\qquad
 R_1^{\rm CBT}(t)=\mathcal N(t),
\end{equation*}
where the last equality follows from equation~\eqref{eq:r1} with
$s(t)=[P_{\rm surv}^{\rm CBT}(t)]^{1/N_s}$ and $s(0)=1$.  Thus the CBT
survival--density equality is a property of the reduced coherent-state model;
comparison with the full-chain MPS remains a nontrivial test of that reduction.
In the continuum long-time limit,
$\sin(E t)/E\rightarrow\pi\delta(E)$ in the distributional sense and
$d\mathcal N/dt$ approaches equation~(\ref{eq:cbtfgr}).  The accessible
window is finite, so we instead apply the same linear estimator used for the
lattice data,
\begin{equation}
 r_{\mathrm{CBT}}(\mathcal I)=
 \frac{\int_{\mathcal I}(t-\bar t)\mathcal N(t)\,dt}
 {\int_{\mathcal I}(t-\bar t)^2\,dt},\qquad
 \delta_{\mathrm{fw}}=
 \frac{r_{\mathrm{CBT}}(\mathcal I)-\Gamma_{\mathrm{CBT,FGR}}}
 {\Gamma_{\mathrm{CBT,FGR}}}.
\label{eq:finitewindowbias}
\end{equation}
Thus $\delta_{\mathrm{fw}}$ is a signed, model-predicted fractional bias,
not a random error and not a criterion for accepting or rejecting MPS evolution.  The finite-window correction
and its associated diagnostic scale are
\begin{equation}
 \rhat_{\mathrm{fw}}=\frac{\rhat_{\mathrm{raw}}}{1+\delta_{\mathrm{fw}}},
 \qquad u_{\mathrm{fw}}=|\rhat_{\mathrm{fw}}-\rhat_{\mathrm{raw}}|.
\label{eq:fwcorrection}
\end{equation}
Flattening the mode
coupling in the diagnostic collapses the reference-point discrepancy, attributing the
dominant offset to the energy dependence of $|\bar\Omega_L|^2$ rather than to
the finite-window fitting procedure.

\subsection{Geometry and uncertainty assessment}

For an OBC total signal, independent bulk nucleation and two endpoints imply
\begin{equation}
 R_{\mathrm{OBC}}^{\mathrm{tot}}(L)
 =L\Gamma_{\mathrm{bulk}}+2\Gamma_{\mathrm{edge}}.
\label{eq:obctotal}
\end{equation}
Dividing by $L$ gives equation~(\ref{eq:obc}); this is why the factor of two
cannot be absorbed silently into an unconstrained $1/L$ term.  For the PBC
density the leading correction is equation~(\ref{eq:pbc}), fitted only when
the complete $L=24,32,48$ size set exists.  A single PBC size supplies a
diagnostic central value but no finite-size extrapolation uncertainty.

The baseline analysis combined regression, window, bond-dimension,
time-step, size, endpoint, readout, and finite-window contributions.  Only
the regression term is a standard error; the others are numerical changes,
half-ranges, extrapolation residuals, or descriptive scales.  In particular,
the included readout dispersion double-counts the algebraically related
$R_3$--$R_4$ correlation/bond pair.  We therefore use this aggregate only to reproduce
the baseline analysis, never as a calibrated confidence interval or as
evidence from branch-screened geometries.  Its full definition, linear-envelope
check, and tables of nearest nonoverlapping intervals are reported in the supplementary material.
\section{Results}

\subsection{Finite-time survival and magnetization benchmark}

The infinite-chain data provide a finite-time cross-level benchmark.  For each
point in the twelve-point parameter set, we compare the L0 full-window survival coefficient
\(\widehat r_{1,g}^{\rm full}\) with the L1 CBT bubble-density slope
\(\Gamma_{{\rm CBT},g}^{\rm win}\) evaluated on the same physical interval.
The matched ratio
\begin{equation}
 C_g=\frac{\widehat r_{1,g}^{\rm full}}
 {\Gamma_{{\rm CBT},g}^{\rm win}}
\label{eq:matchedconsistency}
\end{equation}
lies between 0.880 and 0.920, with median 0.896.  At the reference point,
\(\widehat r_{1}^{\rm full}=1.25720\times10^{-4}J\) per site and
\(\Gamma_{\rm CBT}^{\rm win}=1.39915\times10^{-4}J\) per site, giving
\(C=0.899\).  After allowing an affine scale, the largest reference-trace
residual is 0.134\% of the $R_1$ signal span.  The corresponding centred
$R^2>0.9999$ is retained only as a descriptive statistic: both curves are
smooth and nearly linear on this interval, so it is not evidence by itself.
The coefficient ratio and signal-relative residual establish a finite-time
perturbative two-kink benchmark.  On the nine-point full-physical-window subset,
the median ratio is 0.899.  At the sole analysis point inside the simultaneous
analytic band, $(h_\perp,S_0)=(0.85,5)$, the survival ratio is 0.897.

The same amplitudes also describe the magnetization response.  Define the
lattice magnetization area $X=(1-m_x/M_{0,g})/2$.  Against the CBT true-vacuum
density $\rho_{\rm CBT}=\sum_n n|a_n(t)|^2$, the time-slope ratios across the twelve-point parameter set
span 0.767--0.931; using the full CBT magnetization operator gives
0.768--0.919.  The reference ratios are 0.875 and 0.879, respectively.  On the
nine-point full-physical-window subset the former ratio spans 0.809--0.931,
with median 0.875.  At $(h_\perp,S_0)=(0.85,5)$, the corresponding
magnetization-area ratio is 0.886.

The fixed-velocity \(R_2\) conversion admits a sharper kinetic-model control.
We plant a bubble generator \(\Gamma\) in the coherent-growth law
\begin{equation}
 X(t)=\Gamma\int_0^t x(s)\,{\rm d}s,\qquad
 x(s)=\frac{2\epsilon(fs)}{f},
\label{eq:coherentgrowthcontrol}
\end{equation}
transform the remaining false-vacuum fraction as
\(-\ln[1-X(t)]\), fit it to \(\beta t^2+\mathrm{const}\) on each
physical window, and apply the fixed-velocity conversion
\(\widehat\Gamma_2=\beta/(2Jh_\perp)\).  Although the reference fit has
centred \(R^2=0.9998\), it returns
\(\widehat\Gamma_2/\Gamma=1.235\).  Across the twelve parameter points the ratio spans
1.138--1.319, with median 1.235.  Thus the fixed-\(t^2\), fixed-velocity
mapping has a 13.8--31.9\% deterministic bias, despite a visually excellent
fit.  This combined control does not assign the bias to a single component of
the transformation or growth law.

\begin{figure}[!htbp]
\centering
\includegraphics[width=0.96\linewidth]{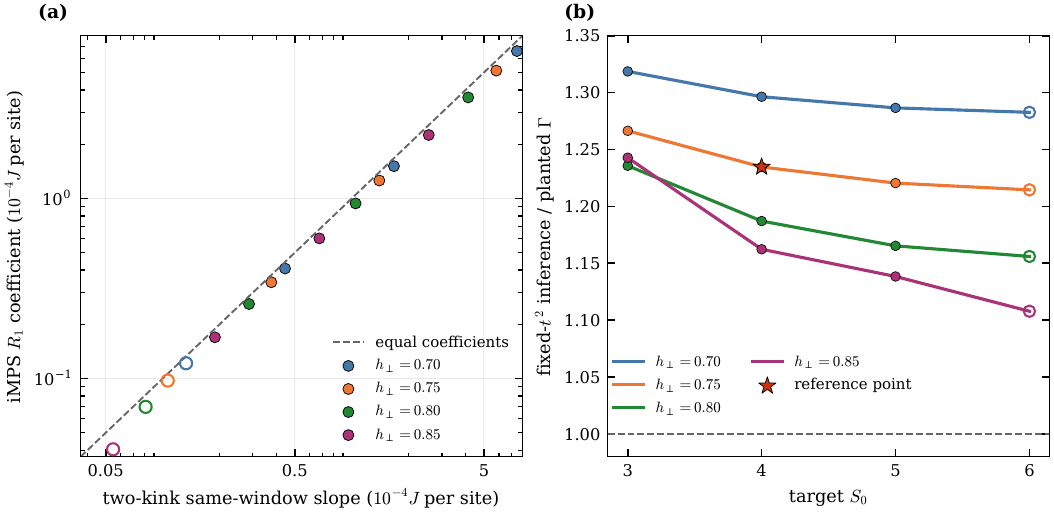}
\caption{Cross-level finite-time comparison for all sixteen principal-grid
points.  Filled markers denote the twelve main analysis points with
$S_0=3,4,5$; open markers denote the four weak-signal points with $S_0=6$, for
which only baseline numerical coverage is available.  (a) The horizontal axis is the coherent
two-kink finite-window slope and the vertical axis is the iMPS $R_1$ coefficient
over the same stored interval; both are in units of $10^{-4}J$ per site and both axes are
logarithmic.  The dashed line denotes equal coefficients, and colour labels the
transverse field $h_\perp$ (dimensionless).  (b) The horizontal axis is the
target action $S_0$ (dimensionless), and the vertical axis is the ratio of the
rate inferred by the fixed-\(t^2\), fixed-velocity \(R_2\)
conversion to the planted coherent-growth generator.  The star marks the
reference point.}
\label{fig:crosslevel}
\end{figure}

The roughly ten-percent survival deficit is systematic rather than scatter.
Replacing the theoretical spontaneous magnetization in the CBT source by the
simulated $M_{0,g}$ changes the matched ratios too little to explain it.  We
therefore report the 0.880--0.920 interval as a matched-window discrepancy
between the full-chain and coherent two-kink calculations, without using it to
recalibrate $\bar\Omega_L$.

\subsection{Bond coherence and observable dependence}

The bond observable behaves differently.  At the last sampled time inside
the reference window, $Jt=10.0$ (the upper window edge is
$Jt_{\rm hi}=10.186$), the MPS value of
$[d_{\rm dw}(t)-d_{\rm dw}(0)]/2$ is
$6.15969\times10^{-3}$, whereas equation~\eqref{eq:bondprojection} gives
$6.83837\times10^{-3}$.  Its vacuum--pair and pair--pair contributions are
$4.97601\times10^{-3}$ and $1.86236\times10^{-3}$, so coherence supplies
72.8\% of the projected signal.  At the reference point, the post-hoc
signal-relative $\chi=64/128$ bond-channel criterion is satisfied; the fixed $R_4$ normalization
cancels from that relative test.

Across the seven projected-bond parameter points satisfying the same
matched-$\chi$ comparison, the endpoint MPS value is 5.2--19.8\% below the
projection (median absolute difference 9.9\%), the MPS-to-projection OLS slope
ratio is 0.276--0.589 (median 0.447), and coherence supplies 60.0--81.5\% of
the projected signal.  The vacuum--pair term therefore rules out a
momentum-diagonal-only description, while the remaining time dependence cannot
be repaired by scalar normalization.  These are observable-level diagnostics:
only two of the seven points satisfy the $R_4$ plateau criterion, five further
points do not satisfy the matched-$\chi$ comparison, and the present data do
not separate higher-kink, inter-bubble, longitudinal-field, initial-state, and
tensor-network contributions to the residual.

\begin{figure}[!t]
\centering
\includegraphics[width=0.98\linewidth]{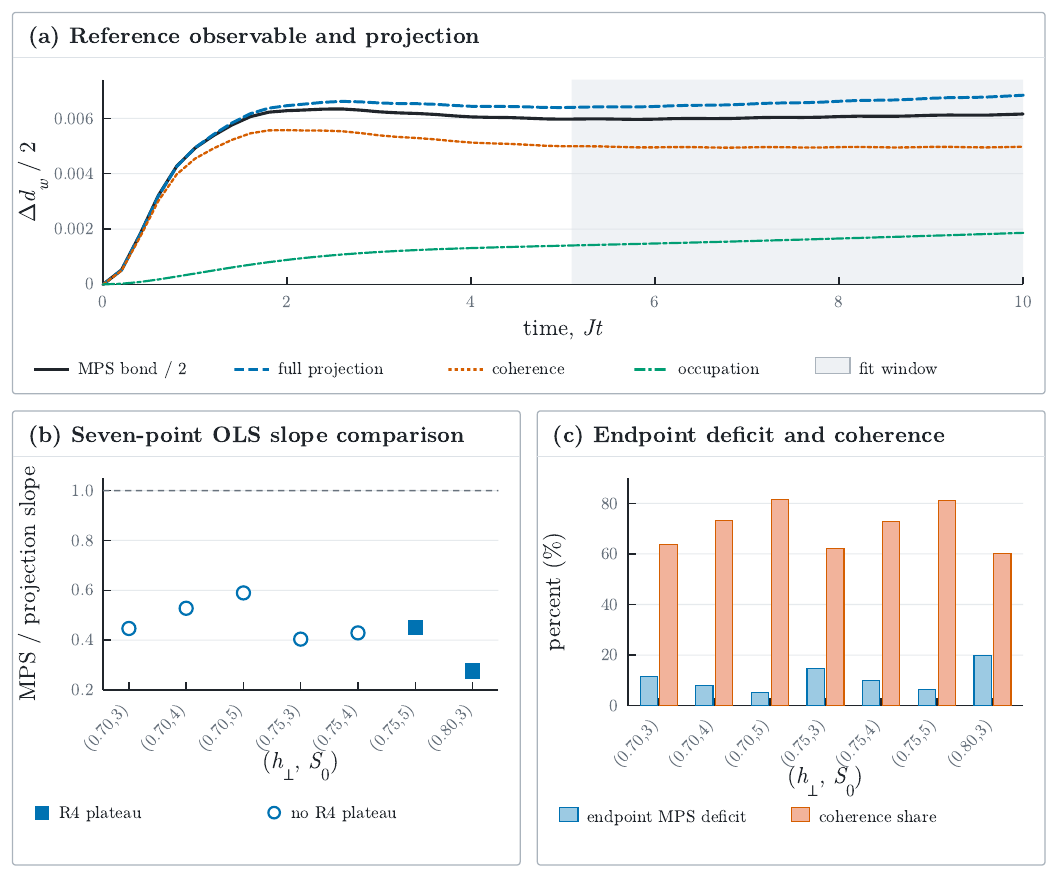}
\caption{Vacuum--two-kink bond-operator comparison.
(a) The horizontal axis is the time $Jt$ and the vertical axis is the
accumulated bond-defect proxy
$\Delta d_{\mathrm{dw}}/2$ (dimensionless).  Curves show the reference MPS proxy, the full
projection, and its vacuum--pair coherence and pair--pair occupation terms;
shading denotes the physical window, and quoted endpoint values use
its last sampled time.  (b) The horizontal labels give the parameter pair
$(h_\perp,S_0)$, whose entries are dimensionless, and the vertical axis is the
MPS-to-projection OLS window-slope ratio for the seven parameter points satisfying the post-hoc matched-$\chi$
bond check.  Filled squares satisfy the stated $R_4$ plateau rule and open
circles do not.  (c) The same horizontal labels are used; the vertical axis
reports, in percent, the endpoint MPS deficit relative to the projection and
the vacuum--pair coherence share.}
\label{fig:bondprojection}
\end{figure}

The quoted values use the source normalization in equation~\eqref{eq:source}.
Under a hypothetical common rescaling
$|\bar\Omega_L|^2\to q_s|\bar\Omega_L|^2$, the vacuum--pair and pair--pair
terms scale as $\sqrt{q_s}$ and $q_s$, respectively.  At the sensitivity point
$q_s=0.896$, chosen only because it approximately sets the median survival
ratio to unity, the coherence range becomes 61.3--82.3\%, the slope-ratio
range becomes 0.309--0.648, and the signed endpoint discrepancy becomes
$-13.4\%$ to $+1.1\%$.  Thus coherence dominance and the late-window slope
mismatch remain, whereas the precise endpoint discrepancy is normalization
sensitive.

The survival transform is a global overlap diagnostic: it measures depletion
of the initial many-body state, and only at leading nontrivial order can it be
identified with a summed excitation probability per site.  The matched
$R_1$--CBT result therefore establishes finite-time consistency between
total-state depletion and the perturbative two-kink spectral structure; it does
not directly count bubbles.  The bond observable probes a distinct local
operator, whose projected expectation contains a vacuum--pair contribution
linear in the two-kink amplitude as well as a quadratic pair--pair term.  Its
unequal late-window slope therefore shows that survival loss cannot be promoted,
without an additional observable-to-rate map, to model-independent local
bubble-number production or a nucleation rate.

\FloatBarrier

\subsection{Finite-size interpretation}

At the reference point,
$(h_\perp,S_0,|h_\parallel|)=(0.75,4,0.05443)$, the admissible interval is
$[5.093,10.186]/J$.  The full-window coefficients are listed in
table~\ref{tab:reference_coefficients}.  Their factor-4.47 spread reflects
readout dependence rather than a confidence interval.

\begin{table}
\caption{Full-window coefficients at the reference point, in $J$ per site.
$R_2$ is the fixed-velocity conversion $\beta/v$; $R_3$ and $R_4$ use the
order-parameter normalization and are not microscopic pair counts.}
\label{tab:reference_coefficients}
\centering
\begin{tabular}{lc}
\toprule
Readout & Full-window coefficient ($J$ per site)\\
\midrule
$R_1$ & $1.25720\times10^{-4}$\\
$R_2$ & $2.17012\times10^{-4}$\\
$R_3$ & $4.85978\times10^{-5}$\\
$R_4$ & $4.85996\times10^{-5}$\\
\bottomrule
\end{tabular}
\end{table}

\begin{table}
\caption{Readout-local finite-time results and finite-size coverage.  ``Positive''
denotes local iMPS plateaux, except for the marked full-window $R_2$
cross-checks.  Finite-state counts use the ten-percent initial-polarization
screen; the five-percent screen gives the same displayed counts.}
\label{tab:funnel}
\centering
\begin{tabular}{lrrrr}
\toprule
Quantity & $R_1$ & $R_2$ & $R_3$ & $R_4$\\
\midrule
Finite-time comparison points & 12 & 12 & 12 & 12\\
Positive local iMPS$^\ast$ & 12 & $12^\ast$ & 6 & 6\\
Initial-state-screened iMPS points & \multicolumn{4}{c}{12}\\
Complete OBC four-size sets & \multicolumn{4}{c}{3}\\
Complete PBC three-size sets & \multicolumn{4}{c}{0}\\
Complete three-geometry points & \multicolumn{4}{c}{0}\\
\bottomrule
\end{tabular}
\end{table}

Table~\ref{tab:funnel} summarizes the finite-size limitation.  All twelve iMPS
points satisfy the initial-state screen, but only three retain a complete OBC
size set and none retains a complete PBC size set in the intended metastable
branch.  At the sole parameter point inside the analytic band, the usable
trajectory ends at $t_\chi$ before completing the physical window.  Thus no
point combines all five analytic scale conditions, a full numerical window,
and a prepared PBC size series, and the available data do not yet support a
bulk-rate extrapolation.

\subsection{Robustness and sensitivity}

At a five-percent signal-relative level, the matched-$\chi$ comparison passes
at 11/12 points for $R_1$, 12/12 for $R_2$, and 7/12 for each of $R_3$ and
$R_4$.  The number of locally stable density-channel slopes varies under
alternative window rules, but none of the tested variants supplies the periodic
size coverage required for a cross-geometry extrapolation.  A synthetic
nucleation-and-growth test recovers its planted rate with a maximum absolute
relative error of 0.620\%, confirming the extraction procedure when its assumed
kinetic structure is present.  Full convergence, cutoff, window-sensitivity,
and recovery results are reported in the supplementary material.
\FloatBarrier
\section{Lattice and continuum semiclassical action benchmark}

\begin{figure}[!t]
\centering
\includegraphics[width=0.96\linewidth]{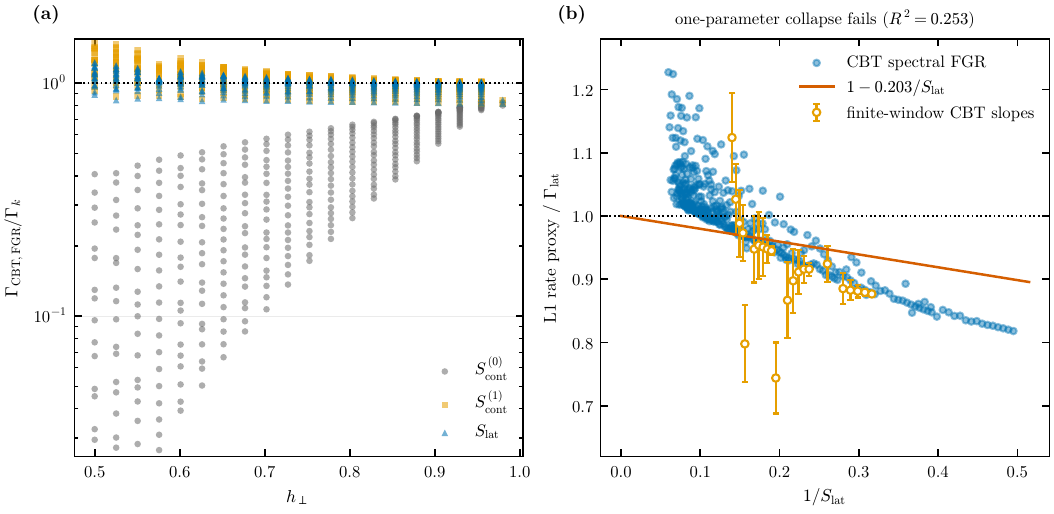}
\caption{L1--L2 fixed-prefactor rate comparison and finite-window L1
diagnostics.  (a) The horizontal axis is the transverse field $h_\perp$
(dimensionless) and the logarithmic vertical axis is the ratio
$\Gamma_{\mathrm{CBT,FGR}}/\Gamma_k$, so the discrepancy uses $\Gamma_k$ in
the denominator through
$\varepsilon_k=|\Gamma_{\mathrm{CBT,FGR}}/\Gamma_k-1|$.
(b) The horizontal axis is the inverse lattice action
$1/S_{\mathrm{lat}}$, and the vertical axis is the L1 rate proxy
divided by $\Gamma_{\mathrm{lat}}$.  Here ``L1 rate proxy'' denotes the
spectral FGR rate for blue points and the finite-window slope at representative
parameter points for orange points.  The constrained line accounts for 25.3\%
of the centred variance ($R^2=0.253$).}
\label{fig:asymptotics}
\end{figure}

We compare the CBT spectral rate with the lattice-resolved and continuum
actions under the common prefactor of equation~\eqref{eq:fixedprefactor}.
Figure~\ref{fig:asymptotics} displays this fixed-prefactor comparison together
with the finite-window L1 diagnostic.

The extended CBT scan yields 382 cutoff-admissible spectra; basis construction
and convergence checks are given in the supplementary material.
Table~\ref{tab:actions} compares equations~\eqref{eq:fixedprefactor} and
\eqref{eq:cbtfgr}.

\begin{table}[!t]
\caption{Fixed-prefactor discrepancies
$\varepsilon_k=|\Gamma_{\mathrm{CBT,FGR}}/\Gamma_k-1|$ for the three actions
over the cutoff-admissible CBT spectra.}
\label{tab:actions}
\centering
\begin{tabular}{lrrr}
\toprule
Action & Median $\varepsilon_k$ & Below 10\% & Below 50\%\\
\midrule
$S_{\mathrm{cont}}^{(0)}$ & 63.01\% & 0.00\% & 34.55\%\\
$S_{\mathrm{cont}}^{(1)}$ & 6.52\% & 64.66\% & 99.48\%\\
$S_{\mathrm{lat}}$ & 4.13\% & 81.41\% & 100.00\%\\
\bottomrule
\end{tabular}
\end{table}

Holding the prefactor fixed isolates the effect of the action choice.  Across
the 382 spectra, 311 lattice-action rate predictions lie within ten percent of
CBT and all lie within fifty percent.  The 4.13\% statistic is the median
$\varepsilon_{\mathrm{lat}}$ of equation~\eqref{eq:epsilon}, not a relative discrepancy between
the actions themselves and not a fit of the full Rutkevich prefactor.

This absolute percentage inherits the shared source normalization.  Under the
hypothetical rescaling $|\bar\Omega_L|^2\to q_s|\bar\Omega_L|^2$,
$C_g\to C_g/q_s$ and
$\Gamma_{\mathrm{CBT,FGR}}\to q_s\Gamma_{\mathrm{CBT,FGR}}$, so
$\varepsilon_k(q_s)=|q_s\Gamma_{\mathrm{CBT,FGR}}/\Gamma_k-1|$.  For
$q_s=0.896$, the continuum-1 and lattice median discrepancies across the 382
spectra become 9.919\% and 9.917\%, respectively, and their counts below ten
percent become 197 and 194.  This counterfactual shows that the absolute 4.13\%
and the fine ordering of those two actions are normalization conditional; it
does not establish $q_s=0.896$ as a physical calibration.  Define the effective
CBT action $S_{\mathrm{eff}}^{\mathrm{CBT}}=-\ln
[2\pi\Gamma_{\mathrm{CBT,FGR}}/f]$ and
$\Delta S_k=S_k-S_{\mathrm{eff}}^{\mathrm{CBT}}$.  After centering each
$\Delta S_k$ distribution, the median absolute deviations are 0.5573, 0.0652,
and 0.0400 for $S_{\mathrm{cont}}^{(0)}$,
$S_{\mathrm{cont}}^{(1)}$, and $S_{\mathrm{lat}}$, respectively.  This
shape comparison is invariant under a global $q_s$ rescaling and retains the
lattice action's advantage in reproducing the parameter dependence.

The constrained $1/S_{\mathrm{lat}}$ residual accounts for 25.3\% of the
centred variance.  The free-intercept diagnostic and the complete 400-cell
classification map are reported in the supplementary material.
\FloatBarrier
\section{Discussion}

\subsection{Physical interpretation and scope}

The same coherent two-kink amplitudes reproduce the finite-time survival and
magnetization dynamics semi-quantitatively, but the inference is observable
dependent.  The survival transform measures total depletion from the initial
many-body state and coincides with a summed excitation probability only
perturbatively.  The bond is a distinct local operator: within the
vacuum--two-kink projection it contains a linear vacuum--pair coherence term
and a quadratic pair--pair term.  Coherence dominance and unequal late-window
slopes therefore show that global survival loss cannot be identified
model-independently with local bubble-number production.  The $R_3$--$R_4$
identity reinforces this interpretation: agreement between those readouts
checks correlation-tail and normalization consistency, but does not provide a
second microscopic bubble count.  Finite-window coefficients nevertheless
remain useful observable-specific diagnostics when their interval and geometry
are stated.

\subsection{Finite-size limitations}

The limitation of the available data arises from two coupled constraints.  The
sole principal-grid point satisfying all five analytic scale conditions,
$(h_\perp,S_0)=(0.85,5)$, reaches the bond-dimension convergence time $t_\chi$
before completing the physical window, placing the analytically compatible
region at the edge of the present tensor-network time reach.  In addition, the
finite-chain data contain no complete PBC size series prepared in the intended
metastable branch.  These constraints, rather than an absence of decay, prevent
the available data from supporting a bulk-rate extrapolation.  Because geometry
and evolution engine also covary at the production scales, a future finite-size
comparison should include a direct cross-algorithm control.

\subsection{Outlook}

A decisive rate identification requires three connected controls: an
observable-to-rate map tested across global and local operators; longer
numerically converged time windows, including higher bond dimension and
adequate correlation range; and finite-size preparation and scaling in the
intended metastable branch, with selecting fields or pins removed before
real-time evolution and common fitting windows across geometries.  Extending
the bond projection to higher-kink or interacting-bubble sectors would test the
origin of its residual dynamics, while a complete prepared PBC size series
would test the thermodynamic extrapolation.  These steps address distinct
operator, time-reach, and finite-size aspects of the inference.

\section{Conclusions}

The same coherent two-kink amplitudes describe finite-time survival and
magnetization semi-quantitatively, whereas the microscopic bond response is
dominated by vacuum--pair coherence and exhibits a late-window mismatch that
cannot be removed by scalar normalization.  This cross-observable contrast
shows that global survival loss cannot be identified directly with local
bubble-number production.

The available finite-size data lack a periodic size series prepared in the
intended metastable branch and therefore do not yet support a bulk-rate
extrapolation.  Within the shared two-kink model, the lattice-resolved action
best reproduces the parameter dependence of the spectral calculation, although
the absolute fixed-prefactor discrepancy remains normalization dependent.
Together, these results separate finite-time reduced-model consistency from
the stronger observable and finite-size evidence required for thermodynamic
rate identification.
\data{The processed data underlying the figures and quantitative results, together with the analysis scripts used in this study, are openly available in Zenodo, DOI: 10.5281/zenodo.21938034. Additional numerical checks and methodological details are provided in the supplementary material.}
\suppdata{Supplementary material provides the readout definitions, action
benchmarks, convergence checks, branch-selection analysis, window diagnostics,
the microscopic bond-operator projection, transparent principal-grid and
claim--sample tables, and extended figures.}

\bibliographystyle{unsrt}
\bibliography{references}

\end{document}